\documentclass[11pt]{article}
\usepackage{graphicx}

\newcommand{\BABARPubYear}    {04}

\input pubboard/babarsym

\newcommand{\UfourS}{\mbox{$\Upsilon(4S)$}}
\def\MeVc2{${\rm MeV}/c^2$}
\def\GeVc2{${\rm GeV}/c^2$}
\def\cont{{q\bar q}}
\def\BR{{\cal BR}}

\def\pipi     {\ensuremath{\pi\pi}}

\def\eg{{\em e.g.}}

\def\mes   {\ensuremath{m_{ES}}}
\def\de   {\ensuremath{\Delta E}}

\def\nno  {\ensuremath{NN}}
\def\coshel {\ensuremath{ cos( \theta_{V}) }}

\def\coshelone {\ensuremath{ cos( \theta_{K^{*+}}) }}
\def\mvone   {\ensuremath{ M_{V1} }}
\def\cosheltwo {\ensuremath{ cos( \theta_{\rho^-}) }}
\def\mvtwo   {\ensuremath{ M_{V2} }}

\def\dz   {\ensuremath{\Delta z}}

\def\qq   {\ensuremath{q \overline{q} }}
\def\bb   {\ensuremath{B \overline{B} }}

\def\ptrue   {\ensuremath{ f_{L} }}

\def\piz {\ensuremath{ \pi^0 }}

\def\de      {\ensuremath{\Delta E}}

\def\borhorho {\ensuremath{B^{0} \rightarrow \rho^+ \rho^- }}
\def\brhorhoo {\ensuremath{B^{+} \rightarrow \rho^+ \rho^0 }}

\def\borhokst {\Bz \rightarrow K^{*+}\rho^-}
\def\borhokstpi {\Bz \rightarrow K^{*+} \rho^-}

\long\def\inst#1{\par\nobreak\kern 4pt\nobreak
    {\it #1}\par\vskip 10pt plus 3pt minus 3pt}

\begin{document}
{\pagestyle{empty}

\begin{flushright}
\babar-CONF-\BABARPubYear/41 \\
SLAC-PUB-10636 \\
August 2004 \\
\end{flushright}

\par\vskip 5cm

\begin{center}
\Large \bf Search for the Decay \boldmath{$\borhokst$} 
\end{center}
\bigskip

\begin{center}
\large The \babar\ Collaboration\\
\mbox{ }\\
\today
\end{center}
\bigskip \bigskip

\begin{center}
\large \bf Abstract
\end{center}
We present the preliminary result of a search for the decay of
$\borhokst$. The data were recorded with the \babar\ detector at the \pep2\ collider and
correspond to 123 million $\bb$ pairs produced in the $e^+e^-$
annihilation through the $\Upsilon(4S)$ resonance.   
We obtain an upper limit on the branching fraction for this decay of 
${\cal B}(\borhokst) <24 \times 10^{-6}$ (90\% C.L.). All results are
preliminary.

\vfill
\vfill
\vfill

\begin{center}

Submitted to the 32$^{\rm nd}$ International Conference on High-Energy Physics, ICHEP 04,\\
16 August---22 August 2004, Beijing, China

\end{center}

\vspace{1.0cm}
\begin{center}
{\em Stanford Linear Accelerator Center, Stanford University, 
Stanford, CA 94309} \\ \vspace{0.1cm}\hrule\vspace{0.1cm}
Work supported in part by Department of Energy contract DE-AC03-76SF00515.
\end{center}

\newpage
} 

\begin{center}
\small

The \babar\ Collaboration,
\bigskip

%
B.~Aubert,
R.~Barate,
D.~Boutigny,
F.~Couderc,
J.-M.~Gaillard,
A.~Hicheur,
Y.~Karyotakis,
J.~P.~Lees,
V.~Tisserand,
A.~Zghiche
\inst{Laboratoire de Physique des Particules, F-74941 Annecy-le-Vieux, France }
% ...

\end{center}\newpage

\section{INTRODUCTION}
\label{sec:Introduction}

\begin{figure}[!ht]
\begin{center}
\vspace{0.2cm}
\rotatebox{-90}{\includegraphics[width=.2\textwidth]{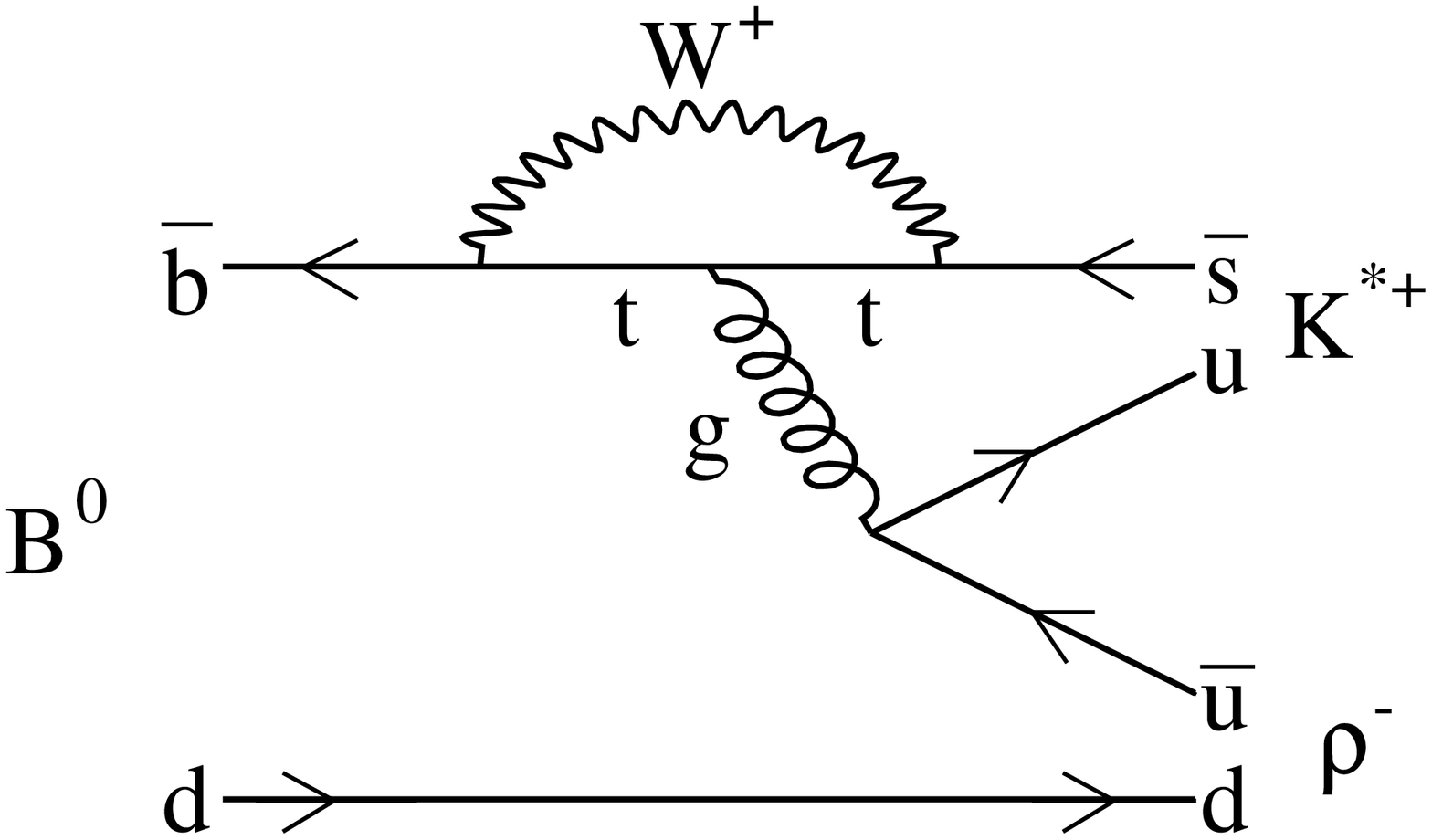}}
\hspace{2cm}
\rotatebox{-90}{\includegraphics[width=.2\textwidth]{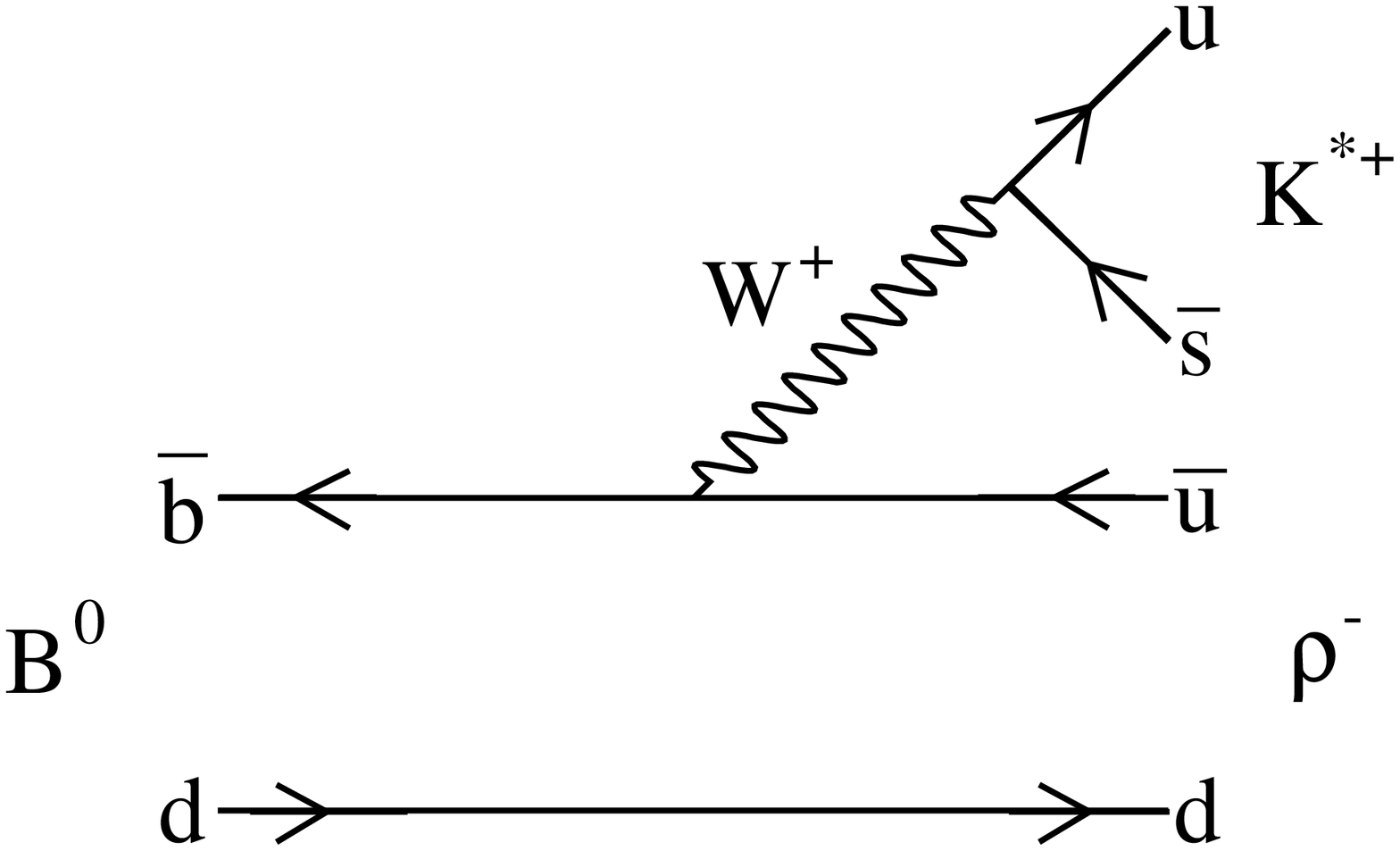}}
\caption{Gluonic penguin and tree diagrams contributing to the process
        $\borhokst$. The penguin contribution coming from the diagrams 
        with $t$ and $c$ quarks in the loop dominates since contributions 
        from process with a u quark is suppressed.
        The tree is a CKM-suppressed diagram.}
\label{fig:feynman}
\end{center}
\end{figure}

The study of the \B\ meson decay into charmless hadronic final states
plays an important role in the understanding of the origin of \CP violation.
The \B decays to two vector particles are of special interest because their
angular distributions reflect both strong- and weak-interaction
dynamics~\cite{palmer}. The decay $\B \rightarrow K^* \rho$ is dominated by 
$b \rightarrow s$ penguin contribution, and the tree-level contribution is CKM
suppressed in the Standard Model (SM) (see Figure~\ref{fig:feynman}). 
The angular correlation measurement is 
particularly sensitive to phenomena beyond the SM, potentially present at either loop- or
tree-level~\cite{kagan}.

\begin{figure}[!ht]
\begin{center}
\vspace{0.2cm}
\resizebox{10cm}{!}{\includegraphics{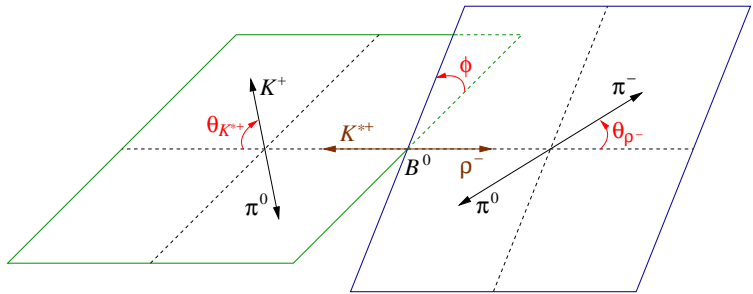}}
\caption{Definition of angles $\theta_{K^{*+}}$, $\theta_{\rho^-}$, and
$\phi$, for the decay $\borhokst$. The $K^+ \pi^0$ ($\pi^-\pi^0$) final
states are shown in the $K^{*+}$ ($\rho^-$) rest frame.}
\label{fig:angle}
\end{center}
\end{figure}

For measurements with limited statistics, we integrate over the
azimuthal angle $\phi$ between the two decay planes shown in Figure~\ref{fig:angle}.
The differential decay rate~\cite{palmer} is 
\begin{equation}
\label{eq:angularcorrelations}
\frac{d^2\Gamma}{\Gamma dcos\theta_{K^{*+}} dcos\theta_{\rho^-} }= \frac{9}{4}\left( \ptrue\
cos^2\theta_{K^{*+}} cos^2\theta_{\rho^-} +
\frac{1}{4}(1-\ptrue) sin^2\theta_{K^{*+}} sin^2\theta_{\rho^-} \right)
\end{equation}
where $\ptrue$ is the longitudinal polarization fraction
component $f_L \equiv \Gamma_L /\Gamma$~\cite{10inbad641,7inbad641}. 
The angles $\theta_{K^{*+}}$ and $\theta_{\rho^-}$ are the helicity angles of $K^{*+}$ and
$\rho^-$, which are defined between the charged $K$($\pi$)
direction and the direction opposite 
the \B in $K^{*+}$($\rho^-$) rest frame as shown in Figure~\ref{fig:angle}. 

In the SM, and assuming naive factorization, 
the polarization is expected to be
proportional to $(1-4\times m^2_\rho/m^2_B)>90\%$~\cite{kagan}, 
which has been verified experimentally in both $\brhorhoo$
and $\borhorho$ decays~\cite{bellerhorhoo, BaBarlbl,rhorho}. 
However, this prediction does not agree with measurements in pure penguin \B decays such as
$B^+ \rightarrow \phi K^{*+}$~\cite{BaBarlbl} and $B^0 \rightarrow \phi K^{*0}$~\cite{new_andrie}  
as shown in Table~\ref{sumfl}. Since the pure penguin \B decay processes are
sensitive to new physics, it is very interesting to look at additonal pure
penguin or penguin dominated \B decays~\cite{kagan}. 

\begin{table}[tbh]
\begin{center}
\caption{Previous measurements for rates and $f_L$ for $B
\rightarrow V_1 V_2$ modes. The first error is statistical and the second error is systematic.}
\vspace{0.5cm}
\begin{tabular}{l|c|c} \hline \hline
 Mode  & ${\cal B}(10^{-6})$  &  $f_L$  \\ \hline
$B^+ \rightarrow \rho^0\rho^+$ 
& $22.5^{+5.7}_{-5.4}\pm 5.8$ & $0.97^{+0.03}_{-0.07}\pm 0.04$ \\
$B^0 \rightarrow \rho^+\rho^-$ 
& $30 \pm 4 \pm 5$ & $0.99 \pm 0.03 \pm 0.04$ \\
$B^+ \rightarrow \rho^0 K^{*+}$ 
& $10.6^{+3.0}_{-2.6}\pm2.4$ & $0.96^{+0.04}_{-0.16}\pm 0.04$ \\
$B^0 \rightarrow \phi  K^{*0}$ 
& $9.2\pm 0.9\pm0.5$ & $0.52 \pm 0.05 \pm 0.02$ \\
$B^+ \rightarrow \phi  K^{*+}$ 
& $12.7^{+2.2}_{-2.0}\pm1.1$ & $0.46 \pm 0.12 \pm 0.03$ \\
\hline \hline
\end{tabular}
\label{sumfl}
\end{center}
\end{table}

In this paper we report a search for the $\borhokst$ decay based on a sample 
of 123 million \BB\ pairs and set a limit on the branching fraction.

\section{THE \babar\ DETECTOR AND DATASET}
\label{sec:babar}

The results presented in this paper are based on data collected
in 1999--2003 with the \babar\ detector~\cite{BABARNIM}
at the PEP-II asymmetric $e^+e^-$ collider
located at the Stanford Linear Accelerator Center (SLAC).  An integrated
luminosity of 113~fb$^{-1}$, corresponding to
123 million \BB\ pairs, was recorded at the $\Upsilon (4S)$
resonance (``on-resonance'') with the center-of-mass(CM) energy ${\ensuremath{\sqrt{s}=10.58\ \gev}}$.
An additional 12~fb$^{-1}$ were taken about 40~MeV below
this energy (``off-resonance'') for the study of continuum backgrounds in
which a light or charm quark pair is produced instead of an \UfourS.

The asymmetric beam configuration in the laboratory frame
provides a boost of $\beta\gamma = 0.56$ to the $\Upsilon(4S)$.
Charged particles are detected and their momenta measured by the
combination of a silicon vertex tracker (SVT), consisting of five layers
of double-sided silicon strip sensors, and a 40-layer central drift chamber,
both operating in the 1.5-T magnetic field of a solenoid.
The tracking system covers 92\% of the solid angle in the CM frame.

Charged-particle identification (PID) is provided by the average
energy loss (\dedx) in the tracking devices  and
by an internally reflecting ring-imaging
Cherenkov detector (DIRC) covering the central region.
A $K/\pi$ separation of better than four standard deviations ($\sigma$)
is achieved for momenta below 3 \gevc , decreasing to 2.5 $\sigma$ at the
highest momenta reached by the $B$ decay final states.
Photons and electrons are detected by a CsI(Tl) electromagnetic calorimeter
(EMC). The EMC provides good energy and angular resolutions for detection of
photons in  the range from 30 \mev\ to 4 \gev. The energy and angular resolutions are
3\% and 4 \mrad\, respectively, for a 1 \gev\ photon.
The flux return for the solenoid is composed of multiple layers of iron
and resistive plate chambers for the identification of muons and long-lived
neutral hadrons.

\section{ANALYSIS METHOD}
\label{sec:Analysis}

Monte Carlo (MC) simulations~\cite{geant4} of the signal decay modes,
continuum and $\bb$ backgrounds are used to establish the event selection
criteria. We reconstruct $\borhokstpi$ candidates from the decay products of
the $K^{*+} \rightarrow K^+ \pi^0$ and $\rho^- \rightarrow \pi^-\pi^0$ 
(the charge conjugate states are implied in this paper). Charged-track candidates are
required to originate from the interaction point: distance of closest
approach to the interaction point less than 10~cm along the beams
direction and less than 1.5~cm in the plane transverse to the beams
direction.
We require that the track from the $\rho^-$ decay has particle identification 
information inconsistent with the electron, kaon, or proton hypotheses; 
and the track from the $K^{*+}$ decay should agree with the kaon hypothesis. 
We reconstruct $\pi^0$ mesons from pairs of photons, where each photon must have
an energy greater than $50\mev$ in the laboratory frame and must exhibit
a lateral profile of energy deposition in the electromagnetic calorimeter
consistent with an electromagnetic shower~\cite{BABARNIM}.
The $\pi^0$ candidates must have a mass that satisfies
$0.11<m(\gamma\gamma)<0.16\gevcc$. 
The mass of the reconstructed $\rho^-$ and $K^{*+}$ candidates must satisfy 
$0.396<m(\pi^{-}\pi^0)<1.146\gevcc$ and $ 0.767<m(K^+\pi^0)<1.017\gevcc$, 
respectively.  Combinatorial backgrounds dominate near $|\coshel|=1$ 
(V denotes $K^{*+}$ or $\rho^-$).
Backgrounds from \B decays, like $\B^0 \rightarrow K^{*+}(892)\pi^-$, with an additional low
energy $\pi^0$ from the rest of the event (ROE), tend to concentrate at
negative values of $\coshel$. The $K^{*+} \to K^+ \pi^0$ and $\rho^- \to \pi^-\pi^0$ 
helicity angles are restricted to the
region $-0.8 < \coshel < 0.98$ to suppress combinatorial background and
reduce acceptance uncertainties due to low-momentum pion reconstruction.

The \B-meson candidates are identified from two nearly
independent kinematic observables~\cite{BABARNIM},
the beam energy-substituted mass $m_{\rm{ES}} \equiv$
$[{ (s/2 + \mathbf{p}_i \cdot \mathbf{p}_B)^2 / E_i^2 -
\mathbf{p}_B^{\,2} }]^{1/2}$ and the energy difference
$\Delta E \equiv (E_i E_B - \mathbf{p}_i
\cdot \mathbf{p}_B - s/2)/\sqrt{s}\,$,
where $(E_i,\mathbf{p}_i)$ is the four-momentum of the 
$e^+e^-$ initial state, and $(E_B,\mathbf{p}_B)$
is the four-momentum of the reconstructed $B$ candidate,
all defined in the laboratory frame.
For signal events, the $m_{\rm{ES}}$ distribution
peaks at the $B$ mass and the $\Delta E$ distribution
peaks near zero. We accept candidates that satisfy 
$5.21 < \mes <5.29~\gevcc$ and $-0.12<\DeltaE<0.15~\gev$. 
The asymmetric \DeltaE window suppresses background from 
higher-multiplicity \B decays. 

In this analysis, $\B^0$ decays to charm modes, such as $\B^0
\rightarrow \bar{D^0} \pi^0$ with $\bar{D^0} \rightarrow K^+\pi^-\pi^0$ have
the same final state as signal.  If the tracks from these \B decays are used
to reconstruct the $K^{*+}$ and $\rho^-$ mesons, these events will 
have peaking $\Delta E $ and $m_{ES}$ distributions under the signal region.
We apply the requirements $|m(K^+\pi^-)-m(\bar{D^0})|>0.02$ GeV$/c^2$ and 
$|m(K^+\pi^-\pi^0)-m(\bar{D^0})|>0.04$ GeV$/c^2$ to
suppress these peaking backgrounds, where $m(\bar{D^0})$ is the nominal mass of $\bar{D^0}$
meson~\cite{ref:pdg2002}. After $\bar{D^0}$ veto 95\%(99\%) of
longitudinal(transverse) signal events are retained. 

Signal candidates may pass the selection even if one or more of the tracks or
$\pi^0$s assigned to the $K^{*+}\rho^-$ state actually comes from the other \B in the event. 
These self-cross-feed (SCF) candidates comprise 37\% (21\%) of the accepted
signal for longitudinal (transverse) signal and are included as signal in the fit. 

Continuum $e^+e^- \to \qq$ ($q = u,d,s,c$) events are the dominant
background. To discriminate signal from continuum we use a neural network 
(\nno) to combine six variables: the Fisher of the Legendre monomials
\cite{pipiBabar}; the sum of transverse momenta in the ROE relative 
to the $z$ axis; the cosine of the angle
between the direction of the \B and the collision axis ($z$) in the
CM frame; the cosine of the angle between the \B thrust
axis and the $z$ axis; the cosine of the angle between the \B-thrust axis and the 
thrust axis of the ROE; the decay angle of one of the $\piz$s (defined in the same 
way as the $K^*$/$\rho$ decay angle, $\theta_V$), randomly selected.
The final sample of signal candidates is selected with a cut on the $\nno$ output that retains 
$\sim 93\%$ ($54\%$) of the signal (continuum).

When multiple \B candidates can be formed, we select the one that 
minimizes the sum of the $\chi^2$ of the reconstructed $\pi^0$ masses from the
nominal $\pi^0$ mass. For those with the same lowest $\chi^2$, we keep the first
one. 

The efficiency of the selection is 6.8\% (13.9\%) for longitudinally
(transversely) polarized signal as determined with MC simulations. 
After the full selections, we obtain 14251 events in the data sample, which are dominated by 
combinatoric backgrounds: roughly $92$\% from \qqbar and $7.7$\% from $\bb$. 

We use MC-simulated events to study the cross-feed
from other $B$~decays.
The charmless modes are grouped into thirteen classes with
similar kinematic and topological properties.
Two additional classes account for the neutral and charged $b \to c$ decays.
For each of the background classes, a component is introduced into the 
likelihood below, with a fixed
number of events.  In the selected $K^{*+}\rho^-$ sample we expect $ 56\pm 27$
charmless background events and $1005$ $b \to c$ events. 

We use an unbinned, extended maximum-likelihood (ML) fit to extract the signal yield. 
The likelihood for each \B candidate $i$ is defined as
\begin{equation}
\label{eq:pdfsum}
{\cal L}_i = e^{-N^{\prime}}\!\prod_{i=1}^{N} \bigg\{ N^{sig}
{\cal P}_{i}^{sig}
+ N^{\cont} {\cal P}_{i}^{\cont}
 + \sum_{j=1} n_j {\cal P}^{\B}_{ji}\bigg\}
\end{equation}
where $N^{\prime}$ is the sum of the signal and continuum yields and the fixed $B$-background
yields, $N^{sig}$ is the number of signal events of type $K^{*+}\rho^-$
in the entire sample.  $N^{\cont}$ is the number of continuum background
events and is floated in the fit. The numbers of events $n_j$ in the \B
background category $j$ are all fixed to their MC expectations.
The probability density function (PDF) ${\cal P}$ is the product of the PDFs of 
seven discriminating variables.
The signal PDF is thus given by ${\cal P}^{sig} =
        {\cal P}(\mes)\cdot
        {\cal P}(\de) \cdot
        {\cal P}(\nno) \cdot
        {\cal P}(\coshelone) \cdot
        {\cal P}(\mvone) \cdot
        {\cal P}(\cosheltwo) 
        \cdot {\cal P}(\mvtwo)$,
where $\mvone$ ($\mvtwo$) is the mass of $K^{*+}$ ($\rho^-$) meson,
and ${\cal P}(\coshelone)$, ${\cal P}(\cosheltwo)$ is the signal helicity
PDF which is expressed as a function of the longitudinal polarization 
(see Eq.~\ref{eq:angularcorrelations}).  The ideal angular distribution
is multiplied by the detector acceptance function.  We obtain the acceptance
function from a simultaneous fit to a sample of MC events with transverse and longitudinal
polarization.
The PDF of the continuum contribution is denoted ${\cal P}^{\cont}$. 
The ${\cal P}^{\B}_{j}$ corresponds to the PDF of 
the \B-background category $j$.
The signal events are decomposed into two parts with distinct distributions:
signal events that are correctly reconstructed and those mis-reconstructed, namely, SCF events.
The SCF fractions for longitudinal and transverse
signal are estimated by MC simulation.
The $\mes$, $\de$, $\nno$, $\coshelone$, $\mvone$, $\cosheltwo$ and
$\mvtwo$ PDFs for signal and $B$ background are taken from the simulation.
The continuum-background $\mes$, $\de$, $\nno$, $\coshelone$ and
$\cosheltwo$ PDF parameters are floated in the fit to data. 
The distributions of the continuum as a function of
$\mvone$ and $\mvtwo$ are described by a non-parametric PDF~\cite{keys} 
derived from \mes\ and \de\ data sidebands.  A total of 12 parameters,
including signal yield and continuum background yield, are varied in
the fit.

\section{SYSTEMATIC UNCERTAINTIES}
\label{sec:Systematics}

The contributions to the systematic error on the signal parameters are
summarized in Table~\ref{tbl:systsummary}.  The uncertainties due to the
signal model are obtained from varying the signal PDF parameters, which are fixed
in the fit, within their estimated errors and assign the effects on the
signal yield as systematic error.  We perform
fits on large MC samples with the measured proportions of $K^{*+}\rho^-$
signal, continuum and \B backgrounds. The bias observed in these tests is
due to imperfections of the PDF model: \eg, unaccounted correlations
between the discriminating variables of the signal and $B$-background
PDFs. The bias is assigned as a systematic uncertainty of 
the fit procedure.  The expected event yields from the \B background modes
are varied according to the uncertainties in the measured or estimated
branching fractions.

\begin{table}[!h]
\begin{center}
\caption{Summary of the systematic uncertainties in the measurement of the 
$\borhokst$ branching fraction.}
\label{tbl:systsummary}
\vspace{0.5cm}
\begin{tabular}{lc} \hline \hline
Source                          & Uncertainty \\\hline
                                & Fit uncertainties (in Events) \\ \hline{}
Signal model                          & $\;_{-1.9}^{+2.3}$    \\
Fit procedure bias                    & $4.2$  \\
\B backgrounds                        & $1.7$ \\ \hline
Total fit error                       & $5.1$ \\\hline
&                                Multiplicative $[\%]$ \\ \hline{}
Track finding                         & 2.4 \\
Neutral correction                   & 10.3 \\
Number of \bb\ pairs                 & 1.1 \\
Particle ID                          & 1.1 \\ \hline
Total multiplicative uncertainties & $10.7\%$ \\ \hline
Non-resonant charmless background     & $\;_{-39.0}^{+0.0}\%$ \\
\hline\hline
\end{tabular}

\end{center}
\end{table}

In this analysis, we do not include a fit component for
other \B decays with the same final-state particles
selected within the $K^*$ or $\rho$ resonance mass window,
such as the non-resonant decays $B^0\to K^+\pi^-\pi^0\pi^0$
, $B^0\to\rho^- K^+ \pi^0$ and $B^0\to K^{*+}\pi^-\pi^0$.
The selection requirements
alone suppress the $B^0\to K^+\pi^-\pi^0\pi^0$ ($B^0\to\rho^- K^+ \pi^0$ and
$B^0\to K^{*+}\pi^-\pi^0$) efficiency by two (one) orders of magnitude
relative to $\borhokst$.
The contribution of these decays to the fit results
is also significantly suppressed
by the mass and helicity-angle information in the fit;
they are examined in the context of mass and helicity-angle
distributions, as discussed below. 

To check the sensitivity of our results to the
presence of non-resonant $B^0\to K^+\pi^-\pi^0\pi^0$
, $B^0\to\rho^- K^+ \pi^0$ and $B^0\to K^{*+}\pi^-\pi^0$ decays, we
explicitly include a fit component for them, assuming
a phase-space decay model. The associated systematic error is estimated by
the difference in the data fit result when the yields of these background
modes are floated or fixed to zero.  We obtain an asymmetric error
of -22 events (-39\%) on the signal yield,
which is systematically overestimated when these non-resonant background
modes are not modeled in the ML fit.  
This systematic error is preliminary estimation, and is presented separately
from the other systematics, with the label "non--resonant".
Interference effects between the resonant and non-resonant
components are ignored in this fit.

The systematic uncertainties in the efficiency
are due to track finding (2.4\% for two tracks), particle identification (1.1\% for two tracks),
and $\pi^0$ reconstruction (10.3\% for two $\pi^0$s).
Smaller systematic uncertainties arise
from event-selection criteria, MC statistics,
and the number of $B$ mesons in the sample.

\section{PHYSICS RESULTS}
\label{sec:Physics}

From the ML fit, we find the signal yield 
$N_{sig} = 58\pm 19(stat)$. The results are summarized in Table~\ref{tab:results}. 
It is checked that we obtain the most conservative estimate for the upper
limit on the branching fraction when 70\% longitudinal polarized signal is
used in the fit. Using the above results, together with the 
selection efficiency, the branching fractions of 
${\cal B}(K^{*+}\rightarrow K^+\pi^0)$,
${\cal B}(\rho^{-}\rightarrow \pi^-\pi^0$), 
${\cal B}(\pi^{0}\rightarrow \gamma\gamma)$,
we obtain a central value of the branching fraction, 
${\cal B}(\borhokst) = [16.3\pm 5.4 (\mathrm{stat}) \pm 2.3(\mathrm{syst})
^{+0.0}_{-6.3}(\mathrm{non-resonant})] \times 10^{-6}$.
The impact of the uncertainties on ${\cal B}(K^{*+}\rightarrow K^+\pi^0)$,
${\cal B}(\rho^{-}\rightarrow \pi^-\pi^0)$, ${\cal B}(\pi^{0}\rightarrow \gamma\gamma)$
is negligible compared to the other systematic errors.
 
In Table~\ref{tab:results}, the statistical error on the branching fraction 
results from the statistical errors on the signal yield.
The systematic uncertainty on the branching fraction results from the propagation of the
systematic uncertainties on signal yield and effective selection efficiency.  Finally, the
non-resonant systematics on signal yield is propagated to the branching fraction.

\begin{figure}[bt]
\begin{center}
\resizebox{12cm}{!}{
        \includegraphics{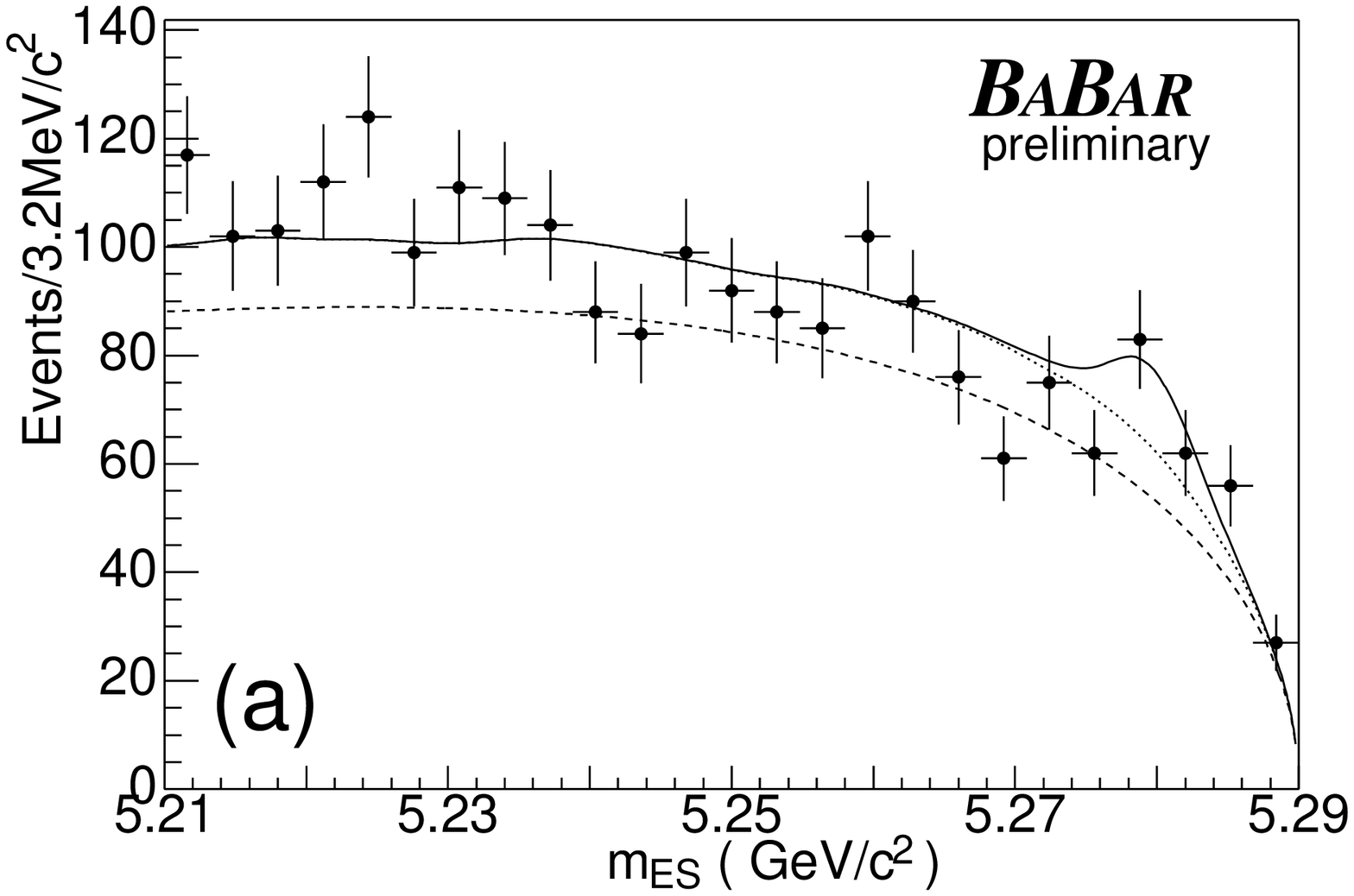}
        \includegraphics{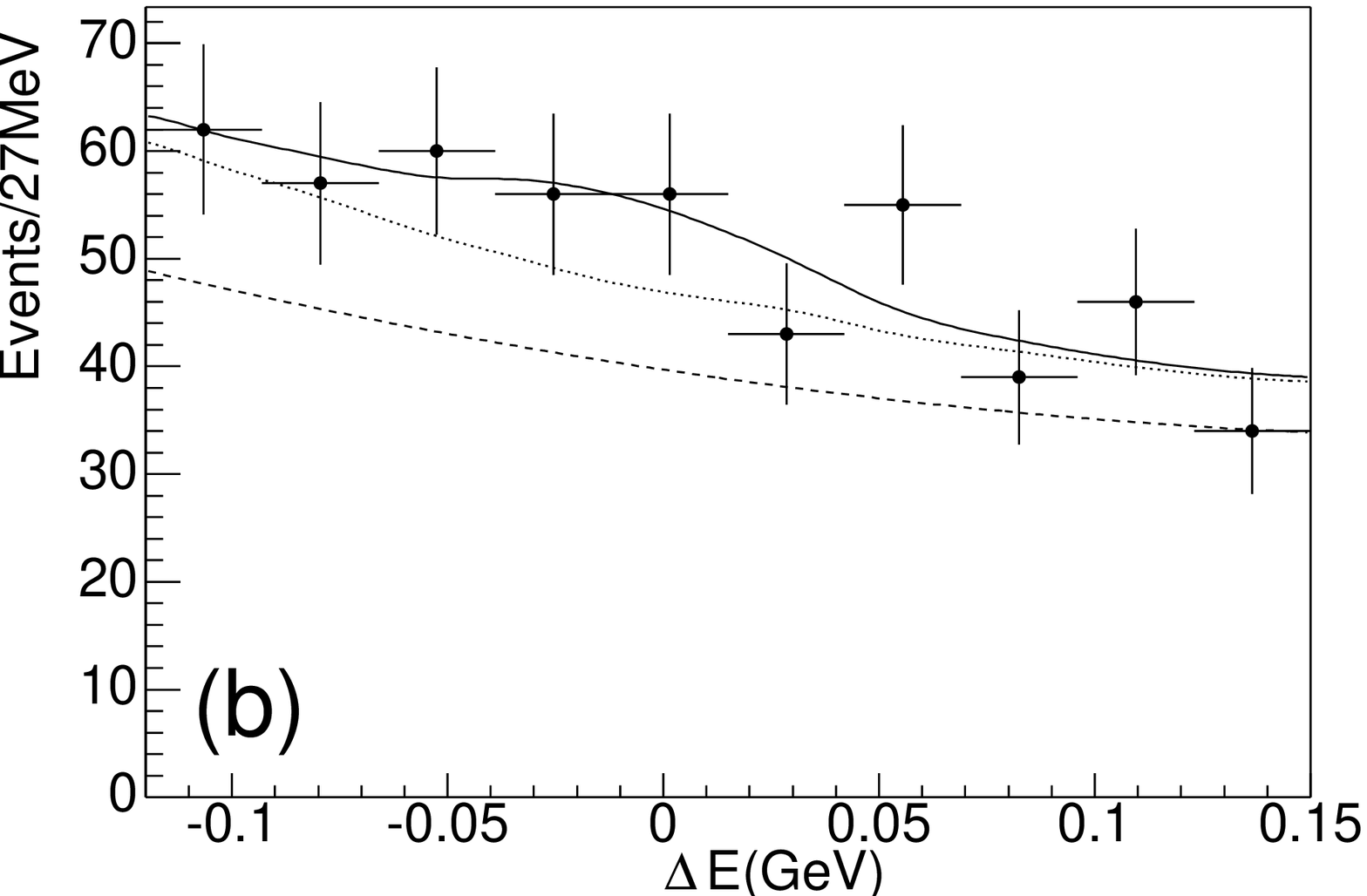}
}
\resizebox{12cm}{!}{
        \includegraphics{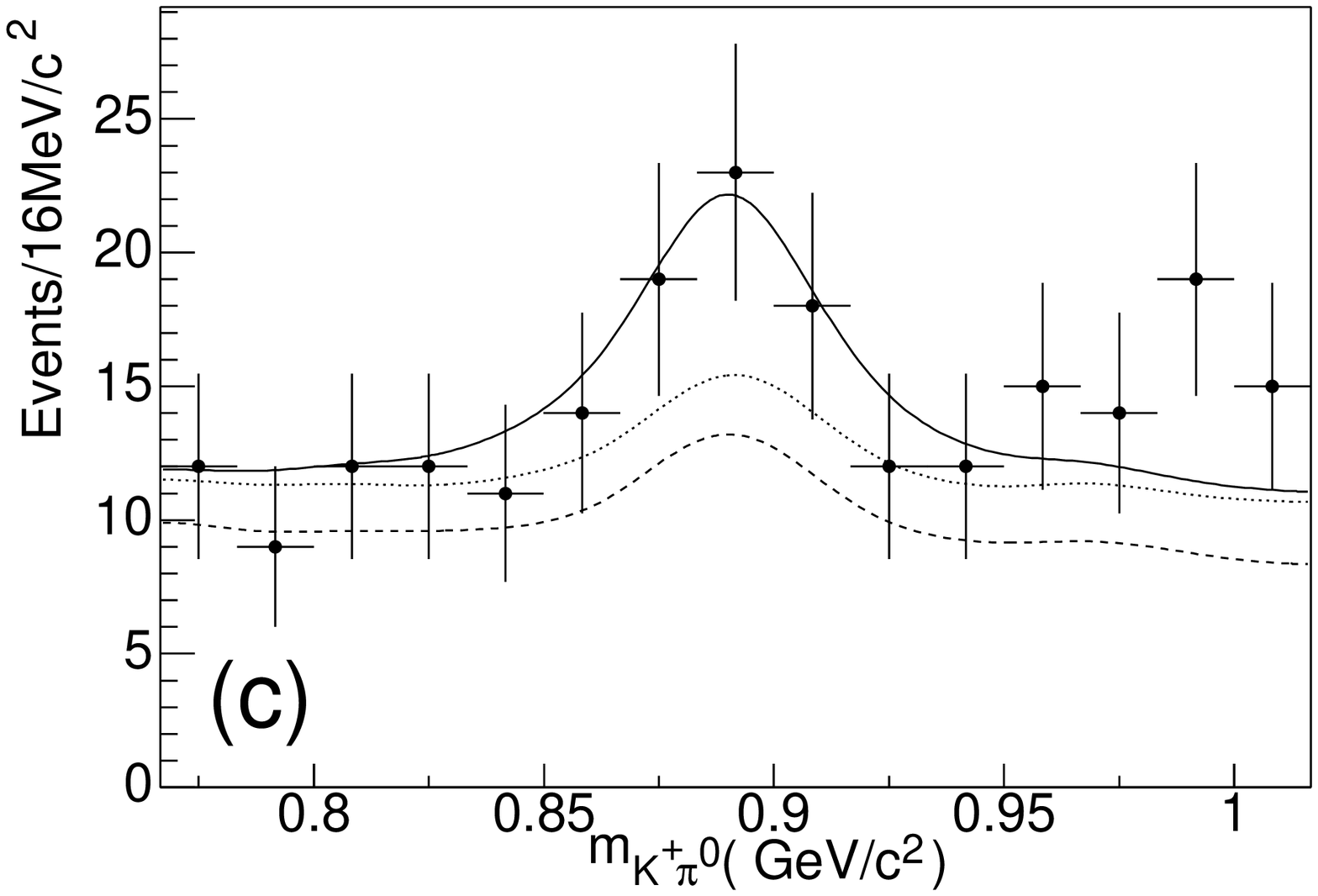}
        \includegraphics{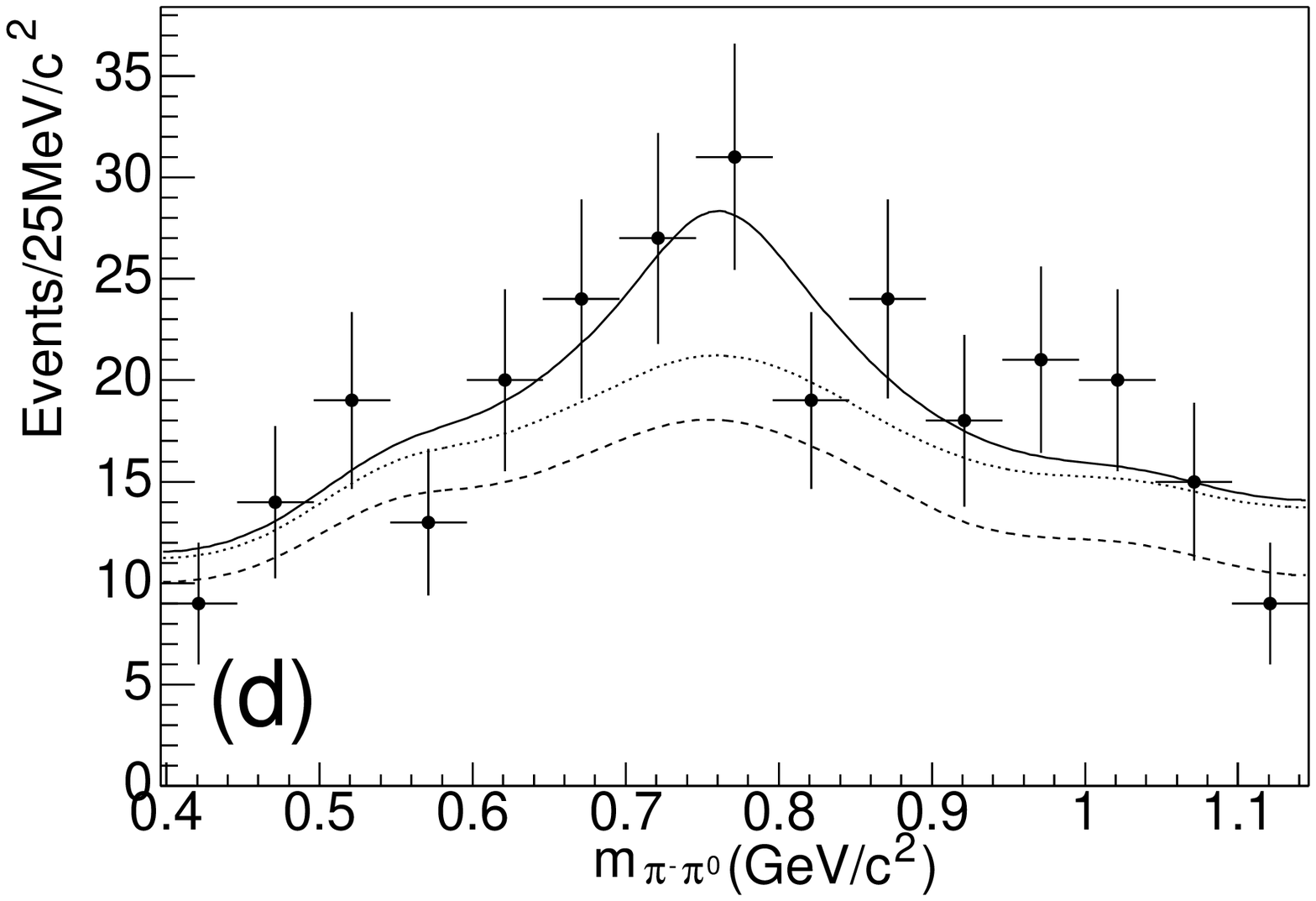}
}
\caption{The (a) \mes, (b) \DeltaE, (c) $K^*$ mass, (d)
$\rho$ mass distribution for signal enriched
samples of the data. The dashed line is the projection of the continuum
background, the dotted line is the projection of the sum of
backgrounds and the solid line is the projection of the fit result. 
}
\label{fig:proj}
\end{center}
\end{figure}

Figure~\ref{fig:proj} shows the result of the fit projected onto the $\mes$,
$\de$, $\mvone$ and $\mvtwo$ observables. The histograms show the data after
a cut on the quantity ${\cal P}_{sig}/({\cal P}_{sig}+{\cal P}_{cont})$ has
been applied, where ${\cal P}_{sig}$ and ${\cal P}_{cont}$ are the
probabilities for a given event to be signal and continuum background, respectively,
and are evaluated using all observables except for the one that is being
plotted. 

Because of the limited statistical significance of the
observed signal, we choose to quote as our preliminary result an upper limit
on the branching fraction. Taking systematic uncertainties into account, an
upper limit of $24 \times 10^{-6}$ at 90\% confidence level (C.L.) is set 
for the $\borhokst$ branching fraction.

\begin{table}[hbtp]
\begin{center}
\caption{Summary of the fit results: Signal yield ($N_{sig}$), effective
selection efficiency ($\varepsilon$),  branching fraction (${\cal B}$), 
upper limit at 90\% confidence level and significance of the measurement,
expressed as number of standard deviation ($\sigma s$). 
The first error corresponds to the statistical uncertainty and the second
one to the systematic uncertainty, and the third
is the systematic uncertainty from non-resonant
contributions.}
\vspace{0.5cm}
\begin{tabular}{c|c} \hline \hline 
 Quantity & Measured Value  \\
\hline
 $N_{sig}$  & $58 \pm 19(stat)$  \\
  $\varepsilon (\%)$ & $8.9 \pm 1.0$  \\
  ${\cal B}(\times{10^{-6}})$ & $16.3 \pm 5.4 (\mathrm{stat})\pm 2.3(\mathrm{syst})^{~+0.0}_{~-6.3}(\mathrm{non-resonant})$  \\
  $U.L.(\times{10^{-6}})$ & $24$ ($22$ statistical only) \\
  Significance ($\sigma$) & $3.2$ ($3.7$ statistical only) \\
\hline \hline
\end{tabular}
\label{tab:results}
\end{center}
\end{table}

\section{SUMMARY}
\label{sec:Summary}

We have searched for the decay $\borhokst$ 
using a maximum likelihood technique in a data sample equivalent 113
fb$^{-1}$ of integrated luminosity.
From a fitted signal yield of $58 \pm 19 \stat$, we obtain
a branching fraction is
$ {\cal B}(\borhokst) = [16.3 \pm 5.4 (\mathrm{stat}) \pm 2.3(\mathrm{syst})
              ^{+0.0}_{-6.3}(\mathrm{non-resonant})] \times 10^{-6} $. 
We get a preliminary upper limit on the $\borhokst$ branching fraction at 90\% C.L. is: 
    \begin{eqnarray*}
      {\cal B}(\borhokst) < 24 \times 10^{-6}.
    \end{eqnarray*}
All results are preliminary.

\section{ACKNOWLEDGMENTS}
\label{sec:Acknowledgments}

We are grateful for the 
extraordinary contributions of our \pep2\ colleagues in
achieving the excellent luminosity and machine conditions
that have made this work possible.
The success of this project also relies critically on the 
expertise and dedication of the computing organizations that 
support \babar.
The collaborating institutions wish to thank 
SLAC for its support and the kind hospitality extended to them. 
This work is supported by the
US Department of Energy
and National Science Foundation, the
Natural Sciences and Engineering Research Council (Canada),
Institute of High Energy Physics (China), the
Commissariat \`a l'Energie Atomique and
Institut National de Physique Nucl\'eaire et de Physique des Particules
(France), the
Bundesministerium f\"ur Bildung und Forschung and
Deutsche Forschungsgemeinschaft
(Germany), the
Istituto Nazionale di Fisica Nucleare (Italy),
the Foundation for Fundamental Research on Matter (The Netherlands),
the Research Council of Norway, the
Ministry of Science and Technology of the Russian Federation, and the
Particle Physics and Astronomy Research Council (United Kingdom). 
Individuals have received support from 
CONACyT (Mexico),
the A. P. Sloan Foundation, 
the Research Corporation,
and the Alexander von Humboldt Foundation.


\begin{thebibliography}{99}

\bibitem{palmer} G. Kramer, W.F.Palmer, {\em Phys. Rev.} {\bf D 45}, 193
                 (1992); H. Y. Cheng, K.C.Yang, {\em Phys. Lett.} {\bf B 511}, 40
                 (2001).
\bibitem{kagan}  A. Kagan, hep-ph/0405134; 
                 P. Colangelo, F. De Fazio and T.N.Pham, hep-ph/0406162.
\bibitem{10inbad641} C.W. Chiang and L. Wolfenstein, {\em Phys. Rev.} {\bf D 61}, 074031 (2000).
\bibitem{7inbad641} G. Kramer, W.F.Palmer and H.Simma, {\em Nucl. Phys.} {\bf B 428}, 77 (1994).
\bibitem{bellerhorhoo} Belle Collaboration, J. Zhang {\em et al.}, {\em Phys.Rev.Lett.} {91}, 221801 (2003).
\bibitem{BaBarlbl} \babar\ Collaboration, B. Aubert {\em et al.}, {\em Phys. Rev. Lett.} {91}, 171802 (2003).
\bibitem{rhorho} \babar\ Collaboration, B. Aubert {\em et al.}, {\em Phys.Rev.} {\bf D 69}, 031102
(2004);\\
\babar\ Collaboration, B. Aubert {\em et al.}, hep-ex/0404029, submitted to {\em Phys. Rev. Lett.}.
\bibitem{new_andrie} \babar\ Collaboration, 
B. Aubert {\em et al.}, hep-ex/0408017, submitted to {\em Phys. Rev. Lett.}.

\bibitem{BABARNIM}
        \babar\ Collaboration, B.\ Aubert {\em et al.},
         Nucl.\ Instrum.\ Methods {\bf A479}, 1-116 (2002).

\bibitem{geant4}
     GEANT4 Collaboration, S. Agostinelli {\em et al.}, 
     Nucl.\ Instrum.\ Methods {\bf A506}, 250 (2003).
\bibitem{pipiBabar}
         \babar\ Collaboration, B.\ Aubert {\em et al.}, {\em Phys. Rev. Lett.} {89}, 281802 (2002).

\bibitem{ref:pdg2002} Particle Data Group, K.~Hagiwara {\em et al.}, Phys.\ Rev.\ {\bf D66}, 010001 (2002).
\bibitem{keys}
        Cranmer, K. S., \cpc{136}, 198 (2001).
\end{thebibliography}
\end{document}